\documentclass[a4paper,fleqn,usenatbib]{mnras}

\usepackage{mathptmx}

\usepackage[T1]{fontenc}
\usepackage{ae,aecompl}

\usepackage{graphicx}	
\usepackage{amsmath}	
\usepackage{amssymb}	

\newcommand{\kms}{km\,s$^{-1}$}

\title[First SiS detection around a Sun-like protostar]{Silicon-bearing molecules in the shock L1157-B1: \\ first detection of SiS around a Sun-like protostar\thanks{Based on observations carried out with the IRAM Plateau de Bure Interferometer. IRAM is supported by INSU/CNRS (France), MPG (Germany) and IGN (Spain).}}

\author[L. Podio et al.]{
L. Podio,$^{1}$\thanks{E-mail: lpodio@arcetri.astro.it}
C. Codella,$^{1}$
B. Lefloch,$^{2}$
N. Balucani,$^{3,1}$
C. Ceccarelli,$^{2,1}$
R. Bachiller,$^{4}$
\newauthor 
M. Benedettini,$^{5}$
J. Cernicharo,$^{6}$
N. Faginas-Lago,$^{3}$
F. Fontani,$^{1}$
A. Gusdorf$^{7}$
\newauthor 
and M. Rosi$^{8}$
\\
$^{1}$INAF - Osservatorio Astrofisico di Arcetri, Largo E. Fermi 5, 50125 Firenze, Italy\\
$^{2}$Univ. Grenoble Alpes, CNRS, IPAG, F-38000 Grenoble, France\\
$^{3}$Dipartimento di Chimica, Biologia e Biotecnologie, Via Elce di Sotto 8, 06123 Perugia, Italy\\
$^{4}$IGN - Observatorio Astron\'omico Nacional, Calle Alfonso XII, 3. 28014 Madrid, Spain\\
$^{5}$INAF, Istituto di Astrofisica e Planetologia Spaziali, via Fosso del Cavaliere 100, 00133 Roma, Italy\\
$^{6}$Grupo de Astrof\'isica Molecular, Instituto de CC. de Materiales de Madrid (ICMM-CSIC), Sor Juana In\'es de la Cruz 3, Cantoblanco,\\
28049, Madrid, Spain\\
$^{7}$LERMA, Observatoire de Paris, \'Ecole normale sup\'erieure, PSL Research University, CNRS, Sorbonne Universit\'es, UPMC Univ. Paris 06,\\
F-75231, Paris, France\\
$^{8}$Dipartimento di Ingegneria Civile ed Ambientale, Via Duranti 93, 06125 Perugia, Italy
}

\date{Accepted XXX. Received YYY; in original form ZZZ}

\pubyear{2015}

\begin{document}
\label{firstpage}
\pagerange{\pageref{firstpage}--\pageref{lastpage}}
\maketitle

\begin{abstract}
The shock L1157-B1 driven by the low-mass protostar L1157-mm is an unique environment to investigate the chemical enrichment due to molecules released from dust grains.
IRAM-30m and Plateau de Bure Interferometer observations allow a census of Si-bearing molecules in L1157-B1. We detect SiO and its isotopologues and, for the first time in a shock, SiS. The strong gradient of the [SiO/SiS] abundance ratio across the shock (from $\ge180$ to $\sim25$) points to a different chemical origin of the two species.
SiO peaks where the jet impacts the cavity walls ([SiO/H$_2$]$\sim 10^{-6}$), indicating that SiO is directly released from grains or rapidly formed from released Si in the strong shock occurring at this location. 
In contrast, SiS is only detected at the head of the cavity opened by previous ejection events ([SiS/H$_2$]$\sim 2 \times 10^{-8}$). This suggests that SiS is not directly released from the grain cores but instead should be formed through slow gas-phase processes using part of the released silicon. 
This finding shows that Si-bearing molecules can be useful to distinguish regions where grains or gas-phase chemistry dominates.
\end{abstract}




\begin{figure}
	\includegraphics[width=6.4cm]{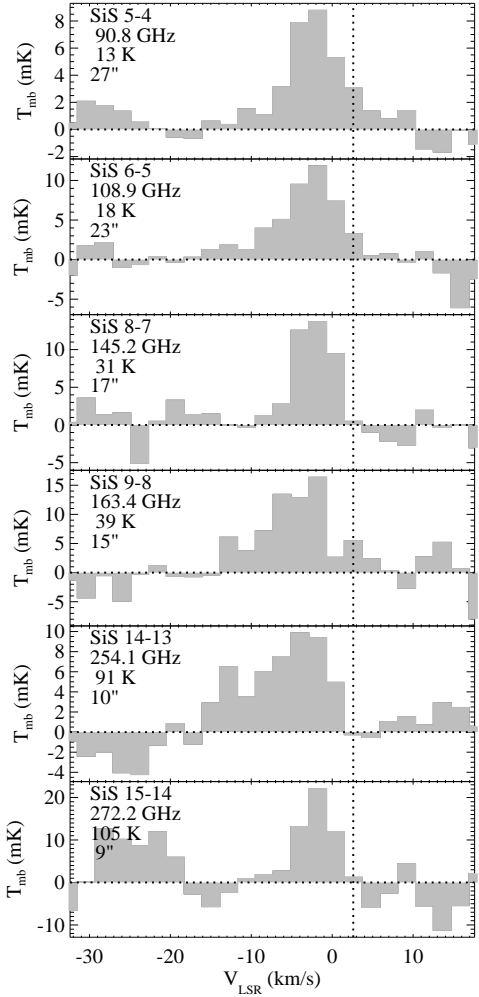}
    \caption{Line profiles of SiS in the L1157-B1 shock. The line intensity is in main beam temperature (T$_{\rm mb}$). The transition, frequency, upper level energy, and HPBW are labeled in the top-left corner. The baseline and the systemic velocity ($+ 2.6$ \kms) are indicated by the horizonthal and vertical dotted lines.}
    \label{fig:spec_sis}
\end{figure}

\begin{figure}
	\includegraphics[width=6.4cm]{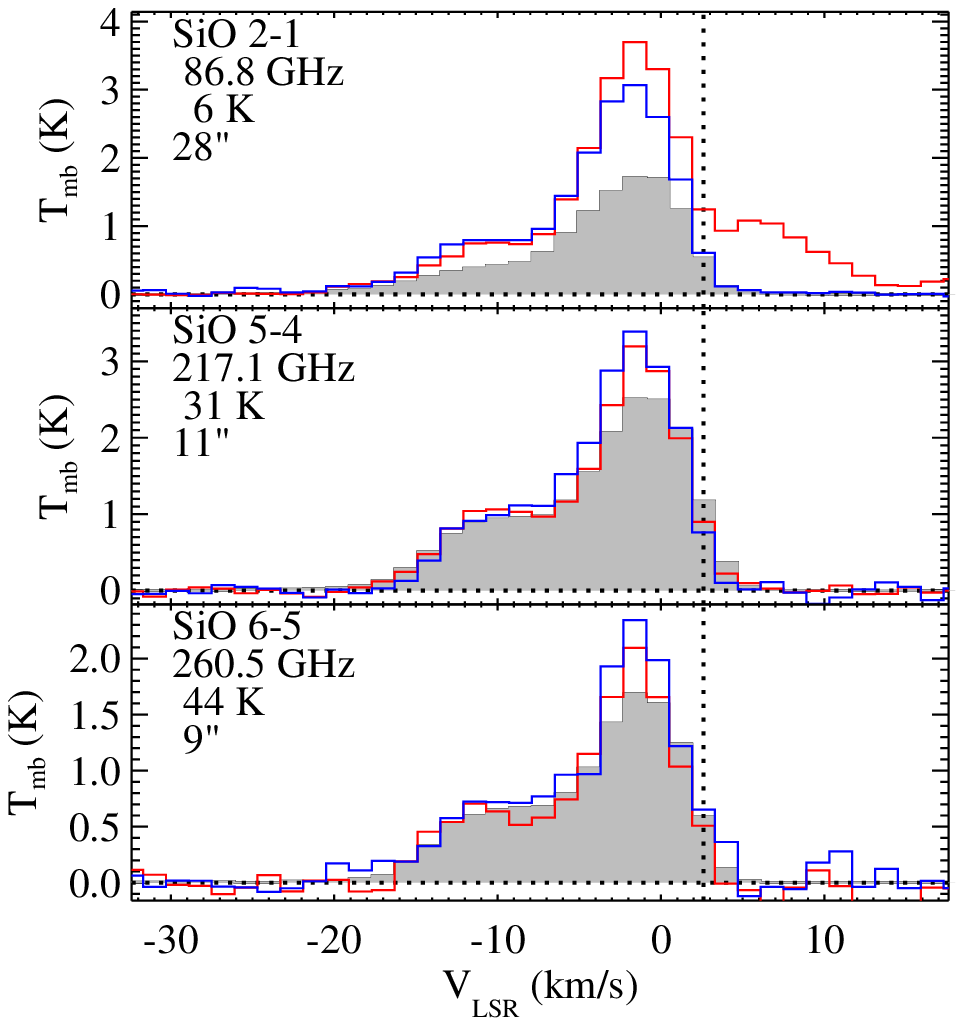}
    \caption{Line profiles of SiO  in the L1157-B1 shock. The  line intensity in main beam temperature (T$_{\rm mb}$) from $^{29}$SiO (red), and $^{30}$SiO (blue) is overplotted on the main isotopologue, $^{28}$SiO (gray), by multiplying for their solar isotopic ratios ([$^{28}$Si/$^{29}$Si]$_{\odot}$ =19.7, [$^{28}$Si/$^{30}$Si]$_{\odot}$ = 29.8, \citealt{anders89}). The transition, frequency, upper level energy, and HPBW for the main isotopologue are labeled. The baseline and the systemic velocity ($+ 2.6$ \kms) are indicated by the horizonthal and vertical dotted lines. The redshifted component which seems to be associated to $^{29}$SiO $2-1$ is due to blending with the $^{13}$CH$_3$OH 4$_{1,3}$-3$_{-2,2}$ line.}
    \label{fig:spec_sio}
\end{figure}

\begin{figure*}
	\includegraphics[width=15cm]{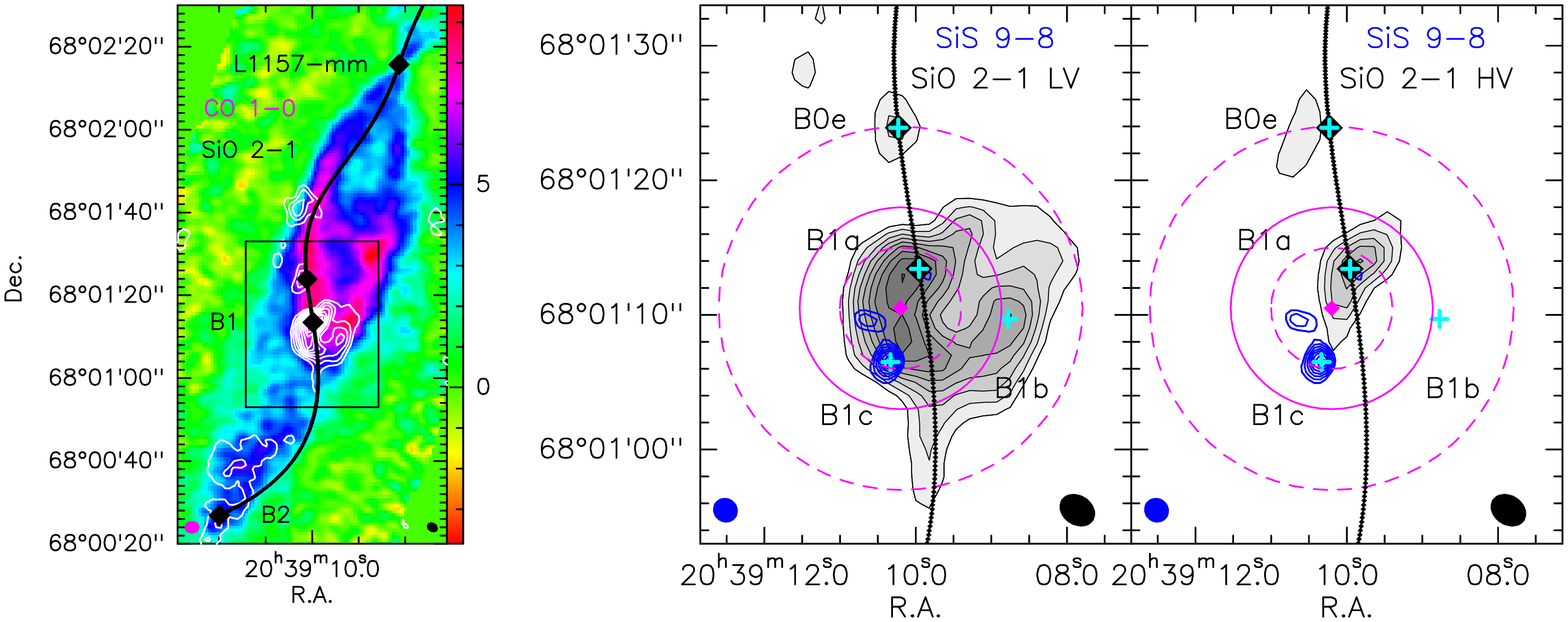}
    \caption{{\it Left panel:} Integrated map of the L1157 blueshifted outflow in CO $1-0$ (colour scale from $-3$ to $9$ Jy\,\kms/beam,  $V_{\rm LSR} = -8/+2.4$ \kms) and SiO $2-1$ (white contours from 1 Jy\,\kms/beam with steps of 0.6 Jy\,\kms/beam, $V_{\rm LSR} =-18/+5$ \kms) \citep{gueth96,gueth98}. The magenta and black ellipses show the HPBW of CO ($3\farcs6 \times 3\farcs0$, PA=90$\degr$) and SiO ($2\farcs8 \times 2\farcs2$, PA=56$\degr$). The black diamonds along the precession pattern modeled by \citet{podio16} (solid black line) indicate the position of the driving source, L1157-mm, and of the spots where the episodic jet impacts on the B2 and B1 cavity walls. {\it Right panels:} Zoom-in on the L1157-B1 cavity. The emission clumps are labeled according to \citet{benedettini07} (cyan crosses). Integrated SiS $9-8$ emission (blue contours, $-10/+2.6$ \kms) is overplotted on SiO $2-1$ (grey scale and black contours) integrated over low velocities (LV: $-8/+5$ \kms) ({\it center}) and over high velocities (HV: $-18/-10$ \kms) ({\it right}). The first contour is at 5$\sigma$ with steps of 1$\sigma$ for SiS and 3$\sigma$ for SiO (1$\sigma = 0.01$ Jy \kms/beam for SiS,  and 0.15 Jy \kms/beam for SiO).  The blue and black ellipses  show the HPBW for SiS ($1\farcs88 \times 1\farcs77$, PA=48$\degr$) and SiO ($2\farcs8 \times 2\farcs2$, PA=56$\degr$), respectively. The magenta diamond and dashed circles show the pointing and minimum/maximum HPBW of the IRAM-30m observations (see Table \ref{tab:lines}). The solid magenta circle show the IRAM-30m HPBW for SiS $9-8$.}
    \label{fig:sis_sio_map}
\end{figure*}

\section{Introduction}

The gravitational infall from which a new star is formed is accompanied by the ejection of highly supersonic jets. The shocks produced when the jet impacts the parental cloud deeply change the chemical composition of the shocked gas due to both gas heating and compression, and dust grain erosion and disruption.
These processes produce a strong enhancement of the abundance of some molecules, which are either released off the dust grains  or formed in the gas \citep[e.g., ][]{bachiller97}.

Some molecules, e.g. H$_2$O, S-bearing molecules, and methanol, are trapped into the icy mantles of dust grains and are released in gas-phase when the dust is heated (e.g. in a hot corino) or eroded in a shock.
At difference, silicon is stored mainly in the core of dust grains  (and in much smaller quantities, $\le10\%$, in the grain mantle or in gas-phase, \citealt{gusdorf08b}) and is likely released in gas-phase only when the grain core is destroyed via gas-grain (sputtering) or grain-grain collisions (shattering and vaporisation) \citep[e.g.][]{gusdorf08a,guillet11}.
Therefore, Si-bearing molecules, such as SiO, are a selective and powerful probe of shocked regions.
However, while SiO has been widely observed in star forming regions and used as a signpost of Class 0 protostars, SiS has only been conclusively observed in the envelope of C-rich stars \citep[e.g., ][]{cernicharo00}, in Sgr B2 \citep{morris75,dickinson81} and in Orion KL \citep{ziurys88,ziurys91,tercero11}. The studies concerning Sgr B2 and Orion suggested that, similarly to SiO, SiS is produced in shocks but the lack of resolved observations prevented to confirm its association to shocks and to investigate its chemical origin.

An interesting target to study the chemistry of shocks at the early protostellar stage is the B1 shock along the L1157 outflow. This is produced by the high-velocity precessing jet driven by the low-mass Class 0 source L1157-mm (L$\sim 3$ L$_{\odot}$, d$\sim 250$ pc) \citep[e.g., ][]{gueth96,podio16}.
Recent observations highlight an unprecedented chemical complexity due to dust grain destruction and show that this region is an ideal laboratory to benchmark chemical models and to investigate the formation routes of ``exotic'' molecular species \citep[e.g., ][]{fontani14b,mendoza14,podio14b,codella15b}.

In this letter we present a survey of Si-bearing molecules in the protostellar shock L1157-B1, based on IRAM-30m and PdBI observations. We report several lines from SiO and the first detection of SiS in a protostellar shock and, more generally, in a low-mass star-forming region. The fractionation and abundance of SiO and SiS are estimated. Then we discuss the chemical origin of SiO and SiS, which may be either directly released from dust grains in the shock or formed through gas-phase processes. This kind of study is a key step towards the understanding the chemical complexity  in the interstellar medium up to the formation of complex organic and pre-biotic molecules.

\section{Observations}

The L1157-B1 protostellar shock ($\alpha_{\rm J2000}$ = 20$^{\rm h}$ 39$^{\rm m}$ 10$\fs$2, $\delta_{\rm J2000}$ = +68$\degr$ 01$\arcmin$ 10$\farcs$5) was observed with the IRAM-30m telescope as part of the ASAI Large Programme\footnote{http://www.oan.es/asai}. The observations were obtained in 2011-2012 with the broad-band EMIR receivers and the FTS spectrometers in their 200 kHz mode (velocity resolutions of 0.17--0.75 \kms).
The IRAM-30m data were processed using GILDAS/CLASS90\footnote{http://www.iram.fr/IRAMFR/GILDAS} software. The spectra were baseline-subtracted and resampled to resolution of $1.4$ \kms\, for SiO and $2.2$ \kms\, for SiS to increase the signal-to-noise. Line intensities were converted from antenna temperature, T$_{\rm a}$, to main-beam brightness temperature, T$_{\rm mb}$, using the main-beam efficiency, $\eta_{\rm mb}$ tabulated on the IRAM website. 

The single-dish data were complemented by high angular resolution observations taken with the IRAM-PdB interferometer (project X058). The SiS $9-8$ line at 163376.78 MHz was observed using the WideX backend, which covers a 4 GHz spectral window with a spectral resolution of $\sim 2$ MHz, i.e. $\sim 3.6$ \kms. The calibration was carried out with GILDAS-CLIC following standard procedures. The clean beam is $1\farcs88 \times 1\farcs77$ (PA=48$\degr$) and the rms noise is $\sim 10$ mJy\,\kms/beam.
The comparison of the SiS $9-8$ spectrum extracted from the PdBI map over a region of 15$''$ with the spectrum obtained with IRAM-30m (HPBW$\sim 15''$) indicates that with PdBI about 75\% of the line flux is recovered (see Fig. \ref{fig:missing_flux} in Appendix \ref{sect:app1}).

\section{Results}


The IRAM-30m spectra show emission from two Si-bearing molecules: 
SiO and its isotopologues $^{29}$SiO and $^{30}$SiO (7 transitions identified from the $2-1$ to the $8-7$, $E_{\rm up}$ from $6$ to $75$ K) and, for the first time in a protostellar shock, SiS (5 lines from the $5-4$  to the $15-14$, $E_{\rm up} \sim 13-105$ K).
The spectra are shown in Figs. \ref{fig:spec_sis} and \ref{fig:spec_sio} and the line properties are summarised in Table \ref{tab:lines} in Appendix \ref{sect:app2}.


The SiS and SiO lines peak at blue-shifted velocity, $\sim -4$~\kms\ with respect to systemic ($V_{\rm sys}=+2.6$ \kms), and show a similar profile at low velocities ($|V - V_{\rm sys}| \le 12$ \kms).
However, the SiS lines are $\sim100-200$ times weaker than the SiO lines arising from upper levels with similar energies. 
Moreover, the SiO lines show a high velocity wing extending up to $-20$~\kms\, likely originating in the strong shock occuring where the jet impacts the cavity walls (see Sect.~\ref{sect:abu}). 

\subsection{Si isotopic ratios and SiO optical depth}
\label{sect:isotopologues}

The ratio between $^{28}$Si and its isotopes $^{29}$Si and $^{30}$Si in the local ISM has been so far poorly constrained.
Observations of $^{29}$SiO and $^{30}$SiO in the Galactic center and in Orion \citep{penzias81,tercero11} indicate an isotopic ratio in agreement with the Solar System ([$^{29}$Si/$^{30}$Si]$_{\odot}$ = 1.5, \citealt{anders89}). 
The secondary-to-primary isotope ratios, [$^{29}$Si/$^{28}$Si] and [$^{30}$Si/$^{28}$Si], are more difficult to estimate because the lowest transitions of $^{28}$SiO are likely optically thick \citep{penzias81}. Observations of $^{28}$SiS and $^{29}$SiS in Orion KL   provided an average value of $^{28}$Si/$^{29}$Si $=26 \pm 10$ \citep{tercero11}, which is consistent with the Solar isotope ratio within the large reported uncertainty ([$^{28}$Si/$^{29}$Si]$_{\odot}$ = 19.7, \citealt{anders89}). Moreover, \citet{monson17} recently developed a method to correct the $^{28}$SiO $1-0$ line for optical depth and showed that the Si isotope ratios are uniform across the Milky Way and shows only a modest increase of the secondary isotopes.
To derive an estimate of the Si isotopic ratios in the L1157-B1 shock we compute the line intensity ratios $^{28}$SiO/$^{29}$SiO  and $^{28}$SiO/$^{30}$SiO in the $6-5$ and $5-4$ lines, which are likely optically thin for all the three isotopologues.
For a given transition the frequencies, upper level energies, and Einstein coefficients for the three isotopologues are comparable, therefore the abundance ratio can be directly derived from the line ratio. 
The $^{28}$SiO/$^{29}$SiO and $^{28}$SiO/$^{30}$SiO ratios in the high velocity wings of the $5-4$ and $6-5$ transitions ($V - V_{\rm sys}= -17/-7$ \kms) are  in excellent agreement with the solar isotopic ratio ($^{28}$SiO/$^{29}$SiO $= 18.8, 19.7$),  while at low velocities the ratios are 10\%-15\% lower ($^{28}$SiO/$^{29}$SiO $= 16.5, 17.3$) (see Fig.~\ref{fig:spec_sio}).
This suggests that opacity effects are responsible for previous estimates of low $^{28}$SiO/$^{29}$SiO and $^{28}$SiO/$^{30}$SiO ratios relative to solar \citep{penzias81} and that Galactic chemical evolution has apparently not led to significant changes of the Si isotope ratios in the past 4.6 billion years in our region of the Galaxy.
Based on the above analysis we adopt Solar isotopic ratios and estimate the optical depth of the $^{28}$SiO $2-1$ line ($\tau \sim 0.7-1.9$).

\subsection{SiO and SiS: distribution and abundance}
\label{sect:abu}

\begin{figure}
	\includegraphics[width=7.cm]{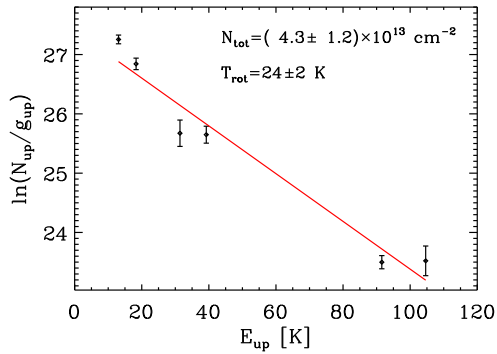}
    \caption{Rotation diagram for SiS.
The derived column densities and rotational temperature are labeled.}
    \label{fig:sis_sio_lte}
\end{figure}

\begin{table}
  \caption[]{\label{tab:abu} Beam-averaged column densities and abundances of SiS and SiO in the clumps B1a and B1c. The last column report the SiO/SiS abundance ratio.}
  \begin{tabular}[h]{cc|cc|cc|c}
    \hline
    \hline
    Clump & $N_{\rm SiS}$                & $N_{\rm SiO}$              & $X_{\rm SiS}$ & $X_{\rm SiO}$ & SiO/SiS \\
               & ($10^{13}$ cm$^{-2}$) & ($10^{14}$ cm$^{-2}$) & ($10^{-8}$) & ($10^{-6}$)    & \\ 
    \hline 
    \hline
    B1a & $\le 0.5$ & $9.3$ & $\le 0.6$ & $1.0$ & $\ge 180$ \\
    B1c & $1.8$       & $4.4$ & $2.0$      & $0.5$ & $25$ \\
   \hline     
  \end{tabular}
\end{table}

The left panel of Fig. \ref{fig:sis_sio_map} shows the structure of the blueshifted outflow driven by the L1157-mm Class 0 protostar in CO $1-0$ and SiO $2-1$ \citep{gueth96,gueth98}.
The high-velocity ($\sim 87$ \kms) precessing jet first observed by \citet{podio16} (black solid line) excavated two main cavities, B1 and B2, located at about $18000$ AU and $33000$ AU distance from L1157-mm. These cavities are well traced by low velocity CO $1-0$ emission \citep{gueth96}. The peaks of SiO $2-1$, instead, probe the regions where the episodic jet impacted the cavity walls at dynamical timescales $t=1800, 1100, 950$ yr (black diamonds). The strong shocks occuring at these locations destroy the dust grains, releasing in gas-phase the silicates in the cores, as well as the molecules frozen on the icy mantles \citep[e.g., ][]{bachiller97,arce08}.
The right panels of Fig. \ref{fig:sis_sio_map}  show a zoom-in of the clumpy structure of the B1 cavity both in SiO $2-1$ (black contours, \citealt{gueth98}) and SiS $9-8$ (blue contours, this work).
The maps indicate that SiS and SiO originate from different regions in the L1157-B1 cavity.
While SiO peaks at the positions where the jet impacted the cavity walls (i.e. B1a and B0e), the bulk of SiS emission is located at the head of the cavity B1c, located $\sim 2000$ AU south of B1a. 
This anti-correlation is even more evident when comparing SiS with the SiO high-velocity emission which is exclusively detected in the B1a and B0e clumps.

In order to investigate the origin of the observed SiS and SiO emission, we derive their column density and abundance. 
First, the rotational diagram of the SiS lines is used to obtain an estimate of the gas temperature and SiS average column density over the cavity. We assume LTE optically thin emission and that the emitting size is the same for all the SiS transitions, i.e. $\sim 3\arcsec$ (see the PdBI map of SiS $9-8$ in Fig.~\ref{fig:sis_sio_map}). We derive $T_{\rm rot} \sim 24\pm2$ K and $N_{\rm SiS} \sim (4 \pm 1) \times 10^{13}$ cm$^{-2}$ (see Fig. \ref{fig:sis_sio_lte}). The derived temperature is in agreement with the value estimated in the cavity from the analysis of other molecules \citep[e.g., ][]{lefloch12,mendoza14,podio14b}.

However, as SiS and SiO probe different regions of L1157-B1, in order to investigate their relative abundance across the shock, we extract the SiS $9-8$ and SiO $2-1$ line intensities from the correponding PdBI datacubes at the position of the B1a and B1c clumps.  The extracted spectra are shown in Fig.~\ref{fig:sis_sio_spec} in Appendix \ref{sect:app3}.
The beam-averaged SiS and SiO column density at the B1a and B1c positions is derived from the SiS $9-8$ and SiO $2-1$ line intensity integrated over their whole velocity profile ($-17/+8$ \kms) assuming optically thin and thermalized emission at $T_{\rm K} \sim T_{\rm rot} = 25$ K, as estimated from the rotational diagram. In the case of SiO the derived column density is corrected for the optical depth across the line profile, i.e. $N_{\rm SiO} = N_{\rm thin} \times \tau / (1-exp^{- \tau})$.
Then, the abundances of SiS and SiO are derived as $X_{\rm X} = X_{\rm CO} \times N_{\rm X} / N_{\rm CO}$, where the column density in the outflow cavity is  $N_{\rm CO} = 9 \times 10^{16}$ cm$^{-2}$ \citep{lefloch12} and the abundance of CO with respect to H$_2$ is assumed to be $X_{\rm CO} = N_{\rm CO} / N_{\rm H_2} = 10^{-4}$. 
Considering the uncertainty on the line fluxes (due to the rms noise as reported in Tab. \ref{tab:lines}, $\le 10\%$, and to calibration, $\sim20\%$), and on the estimated gas temperature in the cavity ($T_{\rm K} \sim 25-80$ K, see \citealt{lefloch12}) the derived column densities and abundances are affected by an error of $\le 30\%$.
The derived SiS and SiO column densities and abundances are summarized in Tab. \ref{tab:abu}.
Interestingly, the relative abundance of SiS with respect to SiO varies by about one order of magnitude across the L1157-B1 shock, i.e. SiO/SiS = 25 in the head of the cavity B1c, and $> 180$ at the jet impact region B1a.

\section{The origin of Si-bearing molecules}

The different abundance and spatial distribution of SiO and SiS strongly suggest a different chemical origin.
There is general consensus that in quiescent clouds most of silicon is trapped into the dust grains cores (only $\le 10\%$ may be in the mantles or in gas-phase in the form of Si or Si$^{+}$, \citealt{gusdorf08b}). However, silicon can be released in shocks in the form of atomic Si and/or SiO due to gas-grain collisions (sputtering) and/or grain-grain collisions (shattering and vaporisation). The latter produces SiO early in the shock while sputtering is only the first step in the production of SiO through the reactions of Si with O$_2$ or OH \citep{gusdorf08a,guillet11}.
It appears then clear that SiO is a natural product of the release from dust grains cores and mantles in the shock. 

On the other hand, the origin of SiS is unclear.
One possible scenario is that SiS is trapped in the core of dust grains and released in gas-phase in the shock. 
Observations of evolved stars show that SiS depletes before SiO. Hence, at difference with SiO, the release of SiS would occur only in strong shocks which are fully destroying the grains.
However, a direct release from dust cores is in contrast with our observations as SiS is not detected at the jet impact site B1a, where we have evidence of a strong shock, but at the head of the cavity in the B1c clump. 

Alternatively, ion-neutral gas-phase processes involving Si$^+$ may convert part of the released silicon into SiS.
This hypothesis is supported by Spitzer observations showing Si$^{+}$ emission in L1157-B1 (Busquet et al., in prep). This suggests that not all the released silicon is converted into SiO and that Si$^{+}$ is easily produced by C$^{+}$ and SiO whose abundance is enhanced in the shock (C$^+$ + SiO $\rightarrow$ Si$^+$ + CO).
According to the existing chemical networks, e.g. KIDA\footnote{http://kida.obs.u-bordeaux1.fr}, the only reaction producing SiS is HSiS$^+$ +  e$^-$ $\rightarrow$ H + SiS, 
where HSiS$^+$ can be formed starting from Si$^{+}$ through the following sets of reactions:
(1) Si$^+$ + OCS $\rightarrow$ SiS$^+$ + CO, SiS$^+$ + H$_2$ $\rightarrow$ HSiS$^+$ + H;
or alternatively (2) Si$^+$ + H$_2$ $\rightarrow$ SiH$_2$$^+$, S + SiH$_2$$^+$ $\rightarrow$ H + HSiS$^+$.
However, laboratory experiments show that SiS$^+$ and Si$^+$ do not react with H$_2$ \citep{wlodek87,wlodek89}.
Hence, the formation of SiS through recombination of HSiS$^{+}$ is unlikely.

Based on the above discussion, we suggest that SiS is likely formed through neutral-neutral gas-phase reactions involving atomic Si or molecules such as SiHn (with n=1, 2, 3, 4), e.g. Si + S/S2/HS and SiHn + S/S2/HS. However, these reactions are not included in the existing chemical networks, hence experimental determinations or accurate theoretical predictions of the reactions barriers and rates are required to verify their efficiency.

In conclusion, according to our observations the L1157-B1 shock consists of two ``chemically'' distinct regions:  the northern region where the jet impacts the cavity walls (B1a clump) is dominated by the release from grains of species formed in the grain cores and mantles, and the older cavity B1c excavated by previous ejection episodes, where new molecules such as SiS are produced by gas-phase processes.

\section{Conclusions}

In this paper we report the first detection of SiS in a low-mass star-forming region and in particular in the L1157-B1 protostellar shock.
Interferometric observations of the shock obtained with the PdBI show for the first time that SiO and SiS have a different spatial distribution. 
We speculate that this is due to the different chemical origins of the two molecules.
SiO peaks where the high-velocity molecular jet reported by \citet{podio16} impacts the cavity walls (the B1a clump) producing a maximum SiO abundance of $\sim 10^{-6}$. Therefore,  SiO is a probe of silicates released from dust grains in shocks.
The SiS emission, instead, is detected only at the head of the cavity in the B1c clump ([SiS/H$_2$] $\sim 2 \times 10^{-8}$).
This points against a direct release of SiS from the grain cores and suggests that SiS should be produced by gas-phase reactions using the released species.
The existing chemical databases indicate only one formation route for SiS, the reaction HSiS$^{+}$ + e$^{-}$ $\rightarrow$ H + SiS.
However, laboratory experiments show that it is not possible to hydrogenate Si$^{+}$ and SiS$^{+}$ to form  HSiS$^{+}$.
Therefore, we propose that SiS is formed by neutral-neutral gas-phase reactions, which are currently not included in the existing chemical networks.
The determination of their reaction rates and their implementation in the chemical network is crucial to understand the chemistry of Si-bearing molecules and will be the subject of a follow-up paper.

\section*{Acknowledgements}
This work was supported by the French program ``Physique et Chimie du Milieu Interstellaire'' (PCMI) funded by CNRS and CNES, and by a grant from LabeX Osug@2020 (Investissements d'avenir - ANR10LABX56). We acknowledge Gemma Busquet for showing us the unpublished Spitzer map of Si$^{+}$ and the referee, Mark Morris, for his helpful suggestions.




\bibliographystyle{mnras}


\newpage
\appendix

\section{Missing flux}
\label{sect:app1}

\begin{figure}
	\includegraphics[width=8.cm]{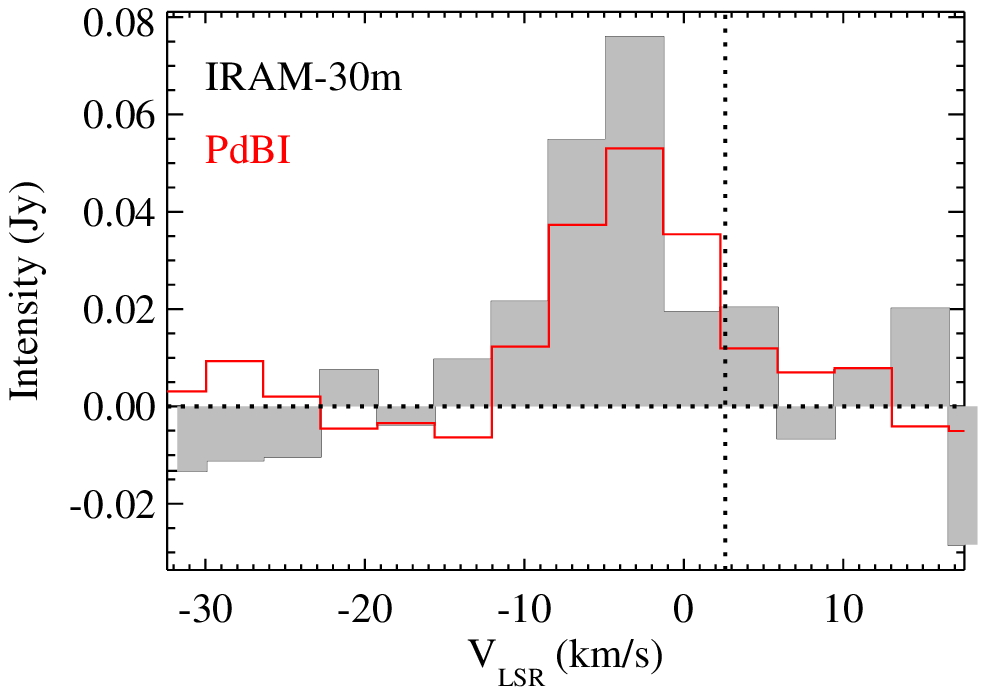}
    \caption{SiS $9-8$ intensity profile extracted from the PdBI datacube integrating on $15"$ area (red line) is overplotted on the IRAM-30m spectrum (HPBW $\sim 15"$) (grey histogram).}
    \label{fig:missing_flux}
\end{figure}

Figure \ref{fig:missing_flux} shows the comparison between the SiS $9-8$ intensity profile extracted from the PdBI datacube over an area of $15"$ and the single-dish IRAM-30m spectrum (HPBW $\sim 15"$). About 75\% $\pm$ 20\% of the flux is recovered by the PdBI observations.

\section{SiO and SiS line properties}
\label{sect:app2}

Table \ref{tab:lines} summarizes the line properties (transition, frequency ($\nu_{\rm 0}$) in MHz, upper level energy (E$_{\rm up}$) in K)  and their observational parameters (telescope half power beam width (HPBW) in arcseconds, peak temperature (T$_{\rm peak}$) in main-beam temperature units, and integrated intensity ($\int {\rm T_{mb} dV}$) in K~\kms).

\begin{table}
\caption[]{\label{tab:lines} Properties of SiO, $^{29}$SiO, $^{30}$SiO, and SiS lines.
}
    \begin{tabular}[h]{cccccc}
\hline
Line & $\nu_{\rm 0}$$^{a}$ & E$_{\rm up}$ & HPBW & T$_{\rm peak}$ & $\int {\rm T_{mb} dV}$$^{b}$ \\
 & (MHz) & (K) & (\arcsec) & (mK) & (K~\kms) \\
\hline
\hline
\multicolumn{6}{c}{SiO} \\
 2-1 &  86846.96 &    6 &   28 & 1715 $\pm$    2 & 16.66 $\pm$  0.01 \\ 
 5-4 & 217104.98 &   31 &   11 & 2514 $\pm$    3 & 26.71 $\pm$  0.02 \\ 
 6-5 & 260518.02 &   44 &    9 & 1688 $\pm$    5 & 17.19 $\pm$  0.03 \\ 
\hline
\multicolumn{6}{c}{$^{29}$SiO} \\
 2-1 &  85759.20 &    6 &   29 &  188 $\pm$    1 &  1.71 $\pm$  0.01 \\ 
 5-4 & 214385.75 &   31 &   11 &  162 $\pm$    3 &  1.44 $\pm$  0.02 \\ 
 6-5 & 257255.22 &   43 &   10 &  106 $\pm$    4 &  0.87 $\pm$  0.03 \\ 
\hline
\multicolumn{6}{c}{$^{30}$SiO} \\
 2-1 &  84746.17 &    6 &   29 &  103 $\pm$    1 &  0.93 $\pm$  0.01 \\ 
 5-4 & 211853.47 &   31 &   12 &  114 $\pm$    3 &  0.99 $\pm$  0.02 \\ 
 6-5 & 254216.66 &   43 &   10 &   79 $\pm$    3 &  0.70 $\pm$  0.02 \\ 
\hline
\multicolumn{6}{c}{SiS} \\
 5-4 &  90771.56 &   13 &   27 &    9 $\pm$    1 & 0.068 $\pm$ 0.005 \\ 
 6-5 & 108924.30 &   18 &   23 &   12 $\pm$    2 & 0.093 $\pm$ 0.009 \\ 
 8-7 & 145227.05 &   31 &   17 &   14 $\pm$    3 & 0.09 $\pm$ 0.02 \\ 
 9-8 & 163376.78 &   39 &   15 &   16 $\pm$    4 & 0.14 $\pm$ 0.02 \\ 
 14-13 & 254103.20 &   91 &   10 &   10 $\pm$    3 & 0.09 $\pm$ 0.01 \\ 
 15-14 & 272243.06 &  105 &    9 &   22 $\pm$    6 & 0.12 $\pm$ 0.03 \\ 
\hline \\
    \end{tabular}
\small
$^{a}$Frequencies are from the CDMS database \citep{muller01}. \\
$^{b}$The line intensity is integrated on the velocity range [$-20, +5$] \kms\, for SiO and [$-10, +2.6$] \kms\, for SiS. 
The reported error is due to the rms noise (the error due to calibration is $\sim20$\%). \\
\end{table}

\section{SiO 2-1 and SiS 9-8 line profiles}
\label{sect:app3}

\begin{figure}
	\includegraphics[width=\columnwidth]{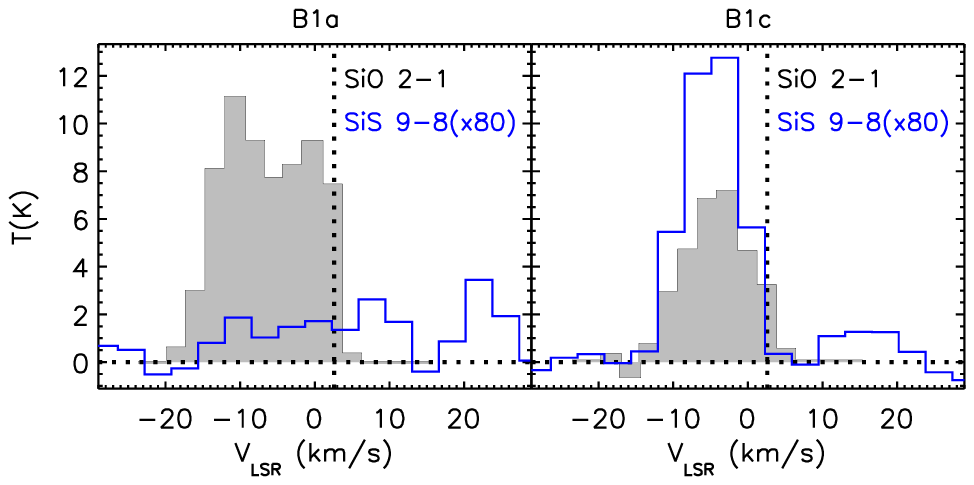}
    \caption{Intensity profiles of SiO $2-1$ (grey histogram) and SiS $9-8$ (blue line, the intensity is multiplied by 80) in the B1a and B1c clumps of the L1157-B1 shock. The baseline and the systemic velocity ($+ 2.6$ \kms) are indicated by the horizonthal and vertical dotted lines.}
    \label{fig:sis_sio_spec}
\end{figure}

Figure \ref{fig:sis_sio_spec} shows the SiS $9-8$ and SiO $2-1$ line intensity extracted from the correponding PdBI datacubes at the position of the B1a and B1c clumps.

\bsp	
\label{lastpage}
\end{document}